\documentclass[aoas,MSNbibl,nameyear,seceqn,dvips]{arximspdf}
\usepackage{dcolumn}
\usepackage{graphicx}

\doi{10.1214/12-AOAS536}
\volume{6}
\issue{3}
\pubyear{2012}
\firstpage{831}
\lastpage{852}

\makeatletter
\newcolumntype{d}[1]{D{.}{.}{#1}}
\newcommand{\calS}{\mathcal{S}}
\newcommand{\calA}{\mathcal{A}}
\newcommand{\calB}{\mathcal{B}}
\newcommand{\calX}{\mathcal{X}}
\newcommand{\indep}{\protect\mathpalette{\protect\independenT}{\perp}}
\def\independenT#1#2{\mathrel{\rlap{$#1#2$}\mkern4.1mu{#1#2}}}
\newcommand{\exc}{\setminus}
\newcommand{\qbox}[1]{\qquad\mbox{#1 }}
\newcommand{\jfor}{\qbox{for}}
\newcommand{\ie}{i.e.}
\makeatother

\begin{document}
\begin{frontmatter}

\title{People born in the Middle East but residing in the Netherlands:
Invariant population size estimates and the role of active and passive
covariates}
\runtitle{Active and passive covariates}

\begin{aug}
\author[A]{\fnms{Peter G. M.} \snm{van der Heijden}\corref{}\ead[label=e1]{p.g.m.vanderheijden@uu.nl}},
\author[B]{\fnms{Joe} \snm{Whittaker}\ead[label=e2]{joe.whittaker@lancaster.ac.uk}},
\author[A]{\fnms{Maarten}~\snm{Cruyff}\ead[label=e3]{m.cruyff@uu.nl}},
\author[C]{\fnms{Bart} \snm{Bakker}\ead[label=e4]{b.bakker@cbs.nl}}
\and
\author[C]{\fnms{Rik} \snm{van der Vliet}\ead[label=e5]{r.vandervliet@cbs.nl}}

\runauthor{P. G. M. van der Heijden et al.}
\affiliation{Utrecht University, Lancaster University, Utrecht
University, Statistics~Netherlands and Statistics Netherlands}
\address[A]{P. G. M. van der Heijden\\
M. Cruyff\\
Department of Methodology\\ \quad and Statistics\\
Utrecht University\\
Postbus 80.140, 3508TC Utrecht\\
The Netherlands\\
\printead{e1}\\
\phantom{\textsc{E-mail}:\ }\printead*{e3}} 
\address[B]{J. Whittaker\\
Department of Mathematics\\ \quad and Statistics\\
Lancaster University\\
Bailrigg\\
Lancaster\\
United Kingdom\\
\printead{e2}}
\address[C]{B. Bakker\\
R. van der Vliet\\
Statistics Netherlands\\
Postbus 24500, 2490HA Den Haag\\
The Netherlands\\
\printead{e4}\\
\phantom{\textsc{E-mail}:\ }\printead*{e5}}
\end{aug}

\received{\smonth{8} \syear{2011}}
\revised{\smonth{1} \syear{2012}}

%
\begin{abstract}
Including covariates in loglinear models of population registers
improves population size estimates for two reasons.
First, it is possible to take heterogeneity of inclusion probabilities
over the levels of a covariate into account; and second, it
allows subdivision of the estimated population by the levels of the
covariates, giving insight into characteristics of individuals
that are not included in any of the registers.
The issue of whether or not marginalizing the full table of registers
by covariates over one or more covariates leaves the estimated
population size estimate invariant is intimately related to
collapsibility of contingency tables [\textit{Biometrika} \textbf{70} (1983) 567--578].
We show that, with information from two registers, population size
invariance is equivalent to the simultaneous collapsibility
of each margin consisting of one register and the covariates.
We give a short path characterization of the loglinear model which
describes when marginalizing over a covariate
leads to different population size estimates. Covariates that are collapsible
are called passive, to distinguish them from covariates that are not collapsible and
are termed active. We make the case that it can be useful to include passive
covariates within the estimation model, because they allow
a finer description of the population in terms of these covariates.
As an example we discuss the estimation of the population size of
people born in the Middle East but residing in the Netherlands.
\end{abstract}

%

\end{frontmatter}

\section{Introduction}\label{sec1}\label{intro}

A well-known technique for estimating the size of a human population
is to find two or more registers of this population, to link the
individuals in the registers and to estimate the number of individuals
that occur in neither of the registers [Fienberg
(\citeyear{FienbergS1972}); Bishop,
Fienberg and Holland (\citeyear{BishopYMMFienbergSEHollandPW1975}); Cormack
(\citeyear{CormackR1989}); International Working Group for Disease
Monitoring and Forecasting, IWGDMF (\citeyear{IWGfDMaF1995})].
For example, with two registers $A$ and $B$, linkage gives a count of
individuals in $A$ but not in $B$, a~count of individuals in $B$ but
not in~$A$, and a count of individuals both in~$A$ and $B$.
The counts form a contingency table denoted by $A\times B$, with the
variable labeled $A$ being short for ``inclusion in register $A$''
taking the levels ``yes'' and ``no,'' and likewise for register $B$.
In this table the cell ``no, no'' has a~zero count by definition, and the
statistical problem is to better estimate this value in the
population.
An improved population size estimate is obtained by adding this
estimated count of missed individuals to the counts of individuals
found in at least one of the registers.

With two registers the usual assumptions under which a population
size estimate is obtained are as follows:
inclusion in register $A$ is independent of inclusion in register
$B$; and
in at least one of the two registers the inclusion probabilities are
homogeneous [see Chao
et al. (\citeyear{ChaoAetal2001}) and Zwane,
van der Pal and van der Heijden (\citeyear{ZwaneEvanderPalKvanderHeijdenPGM2004})].
Interestingly, it is often, but incorrectly, supposed that \textit{both}
inclusion probabilities have to be homogeneous.
Other assumptions are that the population is closed and that it is
possible to link the individuals in registers $A$ and $B$ perfectly.

However, it is generally agreed that these assumptions are unlikely to
hold in human populations.
Three approaches may be adopted to make the impact of possible
violations less severe.
One approach is to include covariates into the model, in particular,
covariates whose levels have heterogeneous inclusion probabilities for
both registers [see Bishop,
Fienberg and Holland (\citeyear{BishopYMMFienbergSEHollandPW1975}); Baker
(\citeyear{BakerSG1990}); compare
Pollock
(\citeyear{PollockKH2002})].
Then loglinear models can be fitted to the higher-way contingency table
of registers $A$ and $B$ and the covariates.
The restrictive independence assumption is replaced by a less
restrictive assumption of independence of $A$ and $B$ conditional on
the covariates; and subpopulation size estimates are derived (one for
every level of the covariates) that add up to a population size
estimate.
Another approach is to include a third register, and to analyze the
three-way contingency table with loglinear models that may include one
or more two-factor interactions, thus getting rid of the independence
assumption.
Here the (less stringent) assumption made is that the three-factor
interaction is absent.
However, including a third register is not always possible, as it is
not available, or because there is no information that makes it
possible to link the individuals in the third register to both the first
and to the second register.
A~third approach makes use of a latent variable to take heterogeneity
of inclusion probabilities into account [see Fienberg, Johnson and Junker (\citeyear{FienbergSJohnsonMJunkerB1999}); Bartolucci
and Forcina (\citeyear{BartolucciFForcinaA2001})].
Of course, these three approaches are not exclusive and may be used
concurrently in one model.

When the approach is adopted to use covariates, the question is which
covariates should be chosen.
In the traditional approach, only covariates that are available in
each of the registers can be chosen.
Recently, Zwane
and van der Heijden (\citeyear{ZwaneEvanderHeijdenPGM2007})
 showed that it is also possible to
use covariates that are not available in each of the registers.
For example, when a covariate is available in register $A$ but not in
$B$, the values of the covariate missed by~$B$ are estimated under a
missing-at-random assumption [Little
and Rubin (\citeyear{LittleRRubinD1987})]; and the subpopulation size estimates
are then derived as a by-product.
Whether or not the covariates are available in each of the registers,
the number of possible loglinear models that can be fit grows rapidly.

In this paper we study the (in)variance of population size estimates
derived from loglinear models that include covariates.
Including covariates in loglinear models of population registers
improves population size estimates for two reasons.
First, it is possible to take heterogeneity of inclusion probabilities
over the levels of a covariate into account; and second, it allows
subdivision of the estimated population by the levels of the
covariates, giving insight into characteristics of individuals that are
not included in any of the registers.
The issue of whether or not marginalizing the full table of registers
by covariates over one or more covariates leaves the estimated
population size estimate invariant is intimately related to
collapsibility of contingency tables.
With information from two registers it is shown that population size
invariance is equivalent to the simultaneous collapsibility of each
margin consisting of one register and the covariates.
Covariates that are collapsible are called passive, to distinguish them
from covariates that are not collapsible and are termed active.
We make the case that it may be useful to include passive covariates
within the estimation model, because they allow a description of the
population in terms of these covariates.
As an example we discuss the estimation of the population size of
people born in the Middle East but residing in the Netherlands.

By focusing on population size estimates, collapsibility in loglinear
models is studied in this paper from a different perspective
than found in Bishop,
Fienberg and Holland (\citeyear{BishopYMMFienbergSEHollandPW1975})
 who are interested
in parametric collapsibility. Our work applies model collapsibility of
Asmussen
and Edwards (\citeyear{AsmussenSEdwardsD1983}), later
discussed by Whittaker
[(\citeyear{WhittakerJ1990}), pages 394--401] and Kim
and Kim (\citeyear{KimSHKimSH2006}),
concerning the commutativity
of model fitting and marginalization. We use model collapsibility in
the context of population size
invariance and show invariance requires model collapsibility of each
margin consisting of one register and the covariates. A novel feature
is to apply collapsibility in the context of a table containing
structural zeros. We give a short
path characterization of the loglinear model which describes when
marginalizing over a covariate leads to different population size
estimates.

The second result can be fruitfully applied in population size
estimation.
In a specific loglinear model, we denote covariates as passive when
they are collapsible and active when they are not collapsible.
In principle, the approach of Zwane
and van der Heijden (\citeyear{ZwaneEvanderHeijdenPGM2007}) permits
the inclusion of many passive covariates in a model; we make a case
for including such passive covariates because they allow the
description of both the observed part as well as the unobserved of the
population in terms of these covariates.

The paper is built up as follows.
In Section~\ref{data} we discuss the data to be analyzed.
These refer to the population of people with Afghan, Iranian and Iraqi
nationality residing in the Netherlands.
In Section~\ref{theory} we discuss theoretical properties of the
loglinear models in the context of population size estimation.
This is discussed in detail for the case of two registers.
We illustrate the two properties of loglinear models using a number of
examples, and then prove the properties using results from graphical
models.
We distinguish the standard situation that every covariate is
available in each of the registers from the situation that there are
one or more covariates that are available in only one of the registers
[Zwane
and van der Heijden (\citeyear{ZwaneEvanderHeijdenPGM2007})].
For completeness we also discuss the situation when three registers
are available and illustrate that the same properties apply.
In Section~\ref{actpass} we develop the notion of active and passive
covariates, and in Section~\ref{sec5} we present an example.
We end with a discussion. In Appendix~\ref{append1} we extend the work of
Asmussen
and Edwards (\citeyear{AsmussenSEdwardsD1983}) to population size invariance.

\section{The population of people with Middle Eastern
nationality staying in the Netherlands}\label{sec2}\label{data}
The preparations for the 2011 round of the Census are in progress
at the time of writing.
More countries now make use of administrative data
(rather than polling) for that purpose.
There are countries who are repeating this method, such as Denmark,
Finland and the Netherlands, and more than ten European countries that
are using administrative data for the first time [Valente
(\citeyear{ValenteP2010})].
The administrative registers are combined by data-linking and
micro-integration to clean and improve consistency.
The outcome of these processes is called a statistical register or
a register for short.

The most important administrative register to be used in the
Netherland Census is an automated system of decentralized (municipal)
population registers (in Dutch, \textit{Gemeentelijke
BasisAdminstratie}, referred to by the abbreviation \textit{GBA}).
This register is used for the definition of the population.
The GBA contains all information on people that are legally allowed to
reside in the Netherlands and are registered as such.
The register is accurate for that part of the population such as
people with the Dutch nationality and foreigners that carry documents
that allow them to be in the Netherlands for work, study, asylum, and
their close relatives.
However, these data do not cover the total population, in particular,
those residing in the Netherlands but who are not allowed to stay
under current Dutch law.
These latter groups are sometimes referred to as undocumented
foreigners or illegal immigrants.

Under Census regulations a quality report is obligatory, and
one of the aspects that needs to be addressed is the undercoverage of
the Census data.
This asks for an estimate of the size of the population that is not
included in the GBA.
In this paper we approach the problem by linking the GBA to another
register and then apply population size estimation methods to arrive
at an estimate of the total population.
Therefore, we implicitly estimate that part of the population not
covered by the GBA.
The second register that we employ is the central Police Recognition
System or HerkenningsDienst Systeem (HKS) that is a collection of
decentralized registration systems kept by 25 separate Dutch police
regions.
In HKS suspects of offences are registered.
Each report of an offence has a suspect identification where, if
possible, information about the suspect is copied from the GBA.
If a suspect does not appear in the GBA, finger prints are taken
so that he or she can be found in the HKS if apprehension at a later
stage occurs.

We test the methodology described in the next sections using
previously collected data of the 15--64 year old age group of people
with Afghan, Iranian or Iraqi nationality.
For the GBA we extract the registered information of 2007.
For HKS we extract information on apprehensions made during 2007.
Table~\ref{comsimple}   illustrates the problem.
For people with Afghan, Iranian or Iraqi nationality $1085 + 26\mbox{,}254 =
27\mbox{,}339$ are registered in the population register GBA; $1085 + 255 =
1340$ are registered in the police register HKS, of whom 255 are
missed by the GBA.
The number of people not in the GBA and not in HKS is to be estimated:
this is the number of people missed by both registers.
This latter estimate plus 255 should be the size of the population
with Afghan, Iranian and Iraqi nationality that do not carry documents
for a legal stay in the Netherlands.
(We ignore the small group of persons who travel on a tourist visa,
and are also not in the GBA and HKS.)
This latter estimate plus ($255 + 1085 + 26\mbox{,}254$) is the size of the
population with Afghan, Iranian or Iraqi nationality that stays in
the Netherlands, either with or without legitimate documents.

\begin{table}
\caption{Linked registers $\mathit{GBA}$ and $\mathit{HKS}$}\label{comsimple}
\begin{tabular}{@{}lcc@{}}
\hline
& \multicolumn{2}{c@{}}{\textbf{HKS}} \\[-5pt]
& \multicolumn{2}{c@{}}{\hrulefill}\\
\textbf{GBA} & \textbf{Included} & \textbf{Not included} \\
\hline
Included & 1085 & 26,254 \\
Not included & \hphantom{1}255 & -- \\
\hline
\end{tabular}
\end{table}

An estimate of the number of people missed by both registers can be
obtained under the assumption that inclusion in GBA is independent of
inclusion in HKS.
In other words, that the odds for in HKS to not in HKS
(1085: 26,254) for the people included in the GBA also holds for
the people not included in the GBA.
The validity of this assumption is difficult to assess.
From a rational choice perspective people without legitimate documents
do their best\vadjust{\goodbreak} to stay out of the hands of the police and so make the
probability of apprehension smaller for those not in the GBA.
On the other hand, people without legitimate documents may be more
involved in activities that lead to a higher probability of apprehension
and so make the probability larger for those not in the GBA.
Both perspectives have face validity but, as far as we know, there is
little empirical evidence to support either. The only relevant work we
found was Hickman
and Suttorp (\citeyear{HickmanLJSuttorpMJ2008}),
who compared the recidivism of deportable and nondeportable
aliens released from the Los Angeles County Jail over a 30-day period in 2002, and found no difference in their
rearrest rates. Yet the relevance of this research for the data at
hand, that discuss people from the Middle-East residing in the
Netherlands, is of course questionable.

With the data at hand, we start from the independence assumption,
but mitigate this by using covariates.
If a covariate is related to inclusion in GBA as well as to inclusion
in HKS but that, conditional on the covariate, inclusion in GBA is
independent of inclusion in HKS, so that ignoring the covariate leads
to dependence between inclusion in GBA and HKS.
For both registers we have gender, age (levels: 15--25, 25--35,
35--50, 50--64) and nationality (levels: Afghan, Iraqi, Iranian).
For GBA we additionally have the covariate marital status (levels:
unmarried, married), and for HKS we have the covariate police
region of apprehension (levels: large urban, not large urban).
We first study theoretical properties for the models employed and then
discuss an analysis of the data.

\section{Theoretical properties of loglinear models}\label{theory}\label{sec3}
\subsection{Two registers, all covariates observed in both
registers}\label{theory2all}\label{sec3.1}
We denote inclusion in the two registers by $A$ and $B$, with levels
$a, b = 1,2$ where level~2 refers to not registered, and we assume
that there are $I$ categorical covariates denoted by $X_i$, where
$i=1,\ldots,I$.
The contingency table classified by variables $A$, $B$ and $X_1$ is
denoted by $A\times B\times X_1$.
We denote hierarchical loglinear models by their highest fitted
margins using the notation of Bishop,
Fienberg and Holland (\citeyear{BishopYMMFienbergSEHollandPW1975}).
For example, in the absence of covariates, the independence model is
denoted by $[A][B]$, and when there is one covariate $X_1$ the model
with $A$ and $B$ conditionally independent given $X_1$ is
$[AX_1][BX_1]$.
In each of the models considered the two-factor interaction between
$A$ and $B$ is absent, as this reflects the (conditional) independence
assumption discussed in the \hyperref[sec1]{Introduction}.

Under the saturated model the number of independent parameters is equal
to the number of observed counts, and the fitted counts are equal to
the observed counts.
The table $A\times B$ has a single structural zero so that the
saturated model is $[A][B]$.
When there are $I$ covariates, the saturated model for the table
$A\times B\times X_1\times \cdots \times X_I$ is
$[AX_1\cdots  X_I][BX_1\cdots  X_I]$, where $A$ and~$B$ are
conditionally independent given the covariates.

We use the following terminology.
We use the word \textit{marginalize} to refer to the contingency table formed
by considering a subset of the original variables.
For example, starting with contingency table $A \times B\times X_1$,
if we marginalize over $X_1$, we obtain the table $A \times B$.
We use the word \textit{collapse} to refer to the situation that when
a table
is marginalized the population size estimate remains invariant.
For example, as we see below, the table $A\times B\times X_1$ is
collapsible over $X_1$ when the loglinear model is $[AX_1][B]$ (or is
$[A][BX_1]$), as the model gives the same population size estimate as
does the $[A][B]$ model for the marginal table $A\times B$.

There are two closely related properties of loglinear models that we
wish to examine:
\begin{longlist}[(2)]
\item[(1)]  There exist loglinear models for which the table is collapsible
over specific covariates.
\item[(2)]  For a given contingency table there exist different loglinear
models that yield identical total population size estimates.
\end{longlist}
The properties are closely related because if Property 2 applies, for
both loglinear models the contingency table to which Property 2 refers
is collapsible over the same covariates. We first illustrate the
properties and then provide an explanation.

%
\begin{figure}[b]

\includegraphics{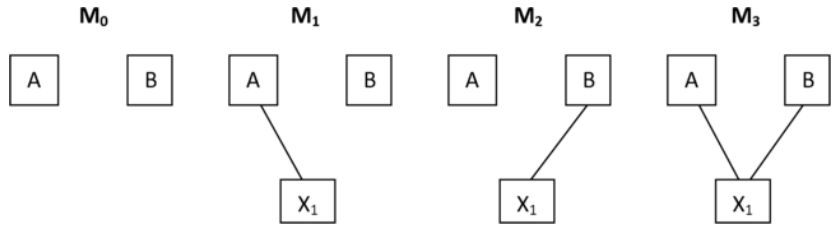}

\caption{Interaction graphs for loglinear models with one
covariate.}\label{Figure1}
\end{figure}

\textit{Example 1.} Assume that there is one covariate $X_1$.
The data are collated in a three-way contingency table
$A\times B\times X_1$.
The total population size estimates under loglinear models
$M_1=[AX_1][B]$ and $M_2=[A][BX_1]$ are equal; this illustrates
Property 2.
Both total population size estimates are equal to the population size
estimate under model $M_0=[A][B]$ in the two-way contingency table $A
\times B$.
Hence, the three-way table is collapsible over $X_1$ and this
illustrates Property 1.
In passing, we note that this result illustrates the second assumption
of population size estimation from two registers discussed in the
\hyperref[sec1]{Introduction}, namely, that the inclusion probabilities only need to be
homogeneous for one of the two registers.
The population size estimate under loglinear model $M_3=[AX_1][BX_1]$
is different from these population size estimates.
See Figure~\ref{Figure1} for interaction graphs of
models $M_0$, $M_1$, $M_2$ and $M_3$.

We present a numerical example in Tables~\ref{models3} and
\ref{counts3}.
Here $A$ refers to inclusion in the official register GBA, $B$ refers
to inclusion in the police register HKS
and the covariate $X_1$ is gender.
See Section~\ref{data} for more details.
We note that, even though the total population size estimates for
models $M_1$ and $M_2$ are equal, estimates of the subpopulations (\ie,
males and females) for $M_1$ are different from those under~$M_2$.

%
\begin{table}
\tabcolsep=0pt
\tablewidth=230pt
\caption{Models fitted to contingency table of variables
$A$ (GBA), $B$~(HKS) and to $A, B$ and $X_1$ (gender), deviances,
degrees of~freedom and estimated numbers missed}\label{models3}
\label{fit}
\begin{tabular*}{220pt}{@{\extracolsep{\fill}}lccc@{}}
\hline
\textbf{Model}&\textbf{Deviance}&\textbf{df}&\textbf{Missed}\\
\hline
$M_0$: $[A][B]$&\hphantom{54}0.0&0&6170.3\\[4pt]
$M_1$: $[AX_1][B]$& 548.5 &1& 6170.3\\
$M_2$: $[A][BX_1]$& \hphantom{54}1.1 & 1& 6170.3\\
$M_3$: $[AX_1][BX_1]$& \hphantom{54}0.0 & 0& 5696.1\\
\hline
\end{tabular*}
\end{table}
%
%
%
\begin{table}[b]
\tabcolsep=0pt
\tablewidth=\textwidth
\caption{Observed and fitted counts for
the three-way table of
$A$ (GBA), $B$ (HKS) and $X_1$ (gender);
for $A$ and $B$ level 1 is present and
for $X_1$ level 1 is male}\label{counts3}
\label{fit}
\begin{tabular*}{\textwidth}{@{\extracolsep{\fill}}lccrrrr@{}}
\hline
$\boldsymbol{A}$&\multicolumn{1}{c}{$\boldsymbol{B}$}&\multicolumn{1}{c}{$\boldsymbol{X_1}$}&\multicolumn{1}{c}{\textbf{obs}}&\multicolumn{1}{c}{$\boldsymbol{M_1}$}&\multicolumn{1}{c}{$\boldsymbol{M_2}$}&\multicolumn{1}{c@{}}{$\boldsymbol{M_3}$}\\
\hline
1& 1& 1& 972 & 629.2 & 976.5 & 972.0\\
2& 1& 1& 234 & 234.0 & 229.5 & 234.0\\
1& 2& 1& 14,883& 15,225.8 & 14,883.0 & 14,883.0\\
2& 2& 1& 0 &5662.2 & 3497.9 & 3582.9\\
1& 1& 2& 113 & 455.8 & 108.5 & 113.0\\
2& 1& 2& 21 & 21.0 & 25.5 & 21.0\\
1& 2& 2& 11,371 & 11,028.2 & 11,371.0 & 11,371.0 \\
2& 2& 2& 0& 508.1 & 2672.5 & 2113.2\\
\hline
\end{tabular*}
\end{table}

\textit{Example 2.} Suppose that there are two covariates, namely, $X_1$
and $X_2$. Table~\ref{models4} presents a fairly comprehensive list of
typical models including the estimated numbers missed and deviances. We
note that models $M_4$, $M_6$ and~$M_{6}'$ have identical total
population size
estimates.
Models $M_5$, $M_8$, $M_9$, $M_{11}$ and
$M_{11}'$ also have identical total
population size estimates.
The remaining models $M_7$, $M_{10}$ and $M_{12}$, $M_{12}'$ and
$M_{12}''$ have different total
population size estimates.

%
%
\begin{table}
\tabcolsep=0pt
\tablewidth=\textwidth
\caption{Models fitted in four-way array of variables $A,
B, X_1$ and $X_2$;
registers $A$ (GBA), $B$~(HKS),
covariates $X_1$ (gender), $X_2$ (age coded in four
levels);
deviances, degrees~of~freedom~and~estimated~numbers~missed}
\label{models4}
\label{fit2}
\begin{tabular*}{\textwidth}{@{\extracolsep{\fill}}lcd{3.1}rr@{}}
\hline
&\textbf{Model}&\multicolumn{1}{c}{\textbf{Deviance}}&\multicolumn{1}{c}{\textbf{df}}&\multicolumn{1}{c@{}}{\textbf{Missed}}\\
\hline
$M_4$& $[AX_1][BX_2]$ & 617.6 & 13 & 6170.3\\
$M_5$& $[AX_1][BX_1][X_2]$ & 228.6 & 15 & 5696.1\\
$M_6$& $[AX_1X_2][B]$ & 718.2 & 7 & 6170.3\\
$M_6'$& $[AX_1][AX_2][X_1X_2][B]$ & 725.6 & 10 & 6170.3\\
$M_7$& $[AX_1][BX_2][X_1X_2]$ & 588.6 & 10 &6179.4\\
$M_8$& $[AX_1][BX_1][BX_2]$ & 69.1 & 12 & 5696.1\\
$M_9$& $[AX_1][BX_1][X_1X_2]$ & 200.2 & 12 &5696.1\\
$M_{10}$& $[AX_1][BX_2][AX_2][BX_1]$ &65.9 & 9 &5837.1\\
$M_{11}$& $[AX_1][BX_1X_2]$ & 4.9 & 6 & 5696.1\\
$M_{11}'$&$[AX_1][BX_1][BX_2][[X_1X_2]$& 34.4 & 9 &5696.1\\
$M_{12}$& $[AX_1X_2][BX_1X_2]$ & 0.0 & 0 &5910.1\\
$M_{12}'$& $[AX_1X_2][BX_1][BX_2]$ & 23.3 & 3 &6257.1\\
$M_{12}''$&$[AX_1][AX_2][BX_1][BX_2][X_1X_2]$ & 31.2 & 6 &5831.4\\
\hline
\end{tabular*}
\end{table}
%
%

%
\begin{figure}[b]

\includegraphics{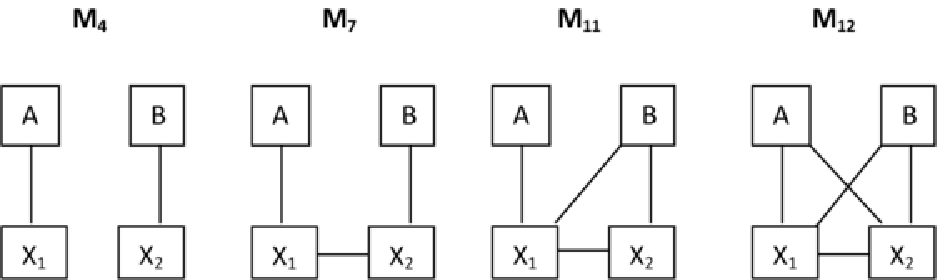}

\caption{Interaction graphs of loglinear models with two
covariates.}\label{Figure2}
\end{figure}

We discuss Properties 1 and 2 together.
We use two notions from graph theory and graphical models, namely, of a
{path} and a {short path} [e.g., see Whittaker
(\citeyear{WhittakerJ1990})].
The two registers $A$ and $B$ are connected by a \textit{path} if
there is a sequence of adjacent edges connecting the variables $A$ and
$B$ in the graph.
A \textit{short path} from $A$ to $B$ is a path that does not contain
a sub-path from $A$ to $B$. Figures
\ref{Figure1} and~\ref{Figure2} illustrate.\vadjust{\goodbreak}
\begin{itemize}
\item In models where $A$ and $B$ are \textit{not} connected, so that
there is no path from $A$ to $B$, the contingency table can be
collapsed over all of the covariates in the graph.
So in Figure~\ref{Figure1} the contingency table
$A\times B\times X_1$ can be collapsed over $X_1$ in model $M_1$ and
in model $M_2$.
This illustrates Property~1 that under models $M_1$ and $M_2$ the
population size estimate is identical to the population size estimate
$M_0$.
In this example this also implies Property~2, that models $M_1$ and
$M_2$ have identical population sizes estimates.
The table $A\times B\times X_1\times X_2$
can be collapsed over both $X_1$ and~$X_2$ in models $M_4$, $M_6$ and
$M_6'$ because $X_1$ and $X_2$ are not on a short path from $A$ to $B$.
In passing, we note this property of model $M_4$ shows that the
inclusion probabilities of $A$ and of $B$ may both be heterogeneous as
long as the sources of
heterogeneity, that is, $X_1$ and $X_2$, are not related.
\item In models with a short path connecting $A$ and $B$, the table is
not collapsible over the covariates in the path.
A simple example is model $M_3$ of Figure~\ref{Figure1}, where the
contingency table $A\times B\times X_1$ cannot be collapsed over
$X_1$. Another simple example is model $M_7$ of Figure~\ref{Figure2},
where the contingency table cannot be collapsed over either $X_1$ or $X_2$.
\item When the covariate $X_2$ is not part of any path from $A$ to $B$
as in models~$M_5$ and $M_8$, then
$A\times B\times X_1\times X_2$ is collapsible over $X_2$,
illustrating Property 1.
Again, for this example, Property 1 implies Property 2, namely, that
these models have identical population size estimates.
\item For model $M_{11}$ of Figure~\ref{Figure2} there are two paths
from $A$ to $B$,
$A-X_1-B$ and $A-X_1-X_2-B$; however, the table is collapsible over
$X_2$, as the second path is not short, containing the unnecessary
detour $X_1-X_2-B$.
\item The other models have no covariates over which the contingency
table can be collapsed. For example, in
model $M_{12}$ of Figure~\ref{Figure2}, and its reduced versions
$M_{12}'$ and $M_{12}''$,
there are again two short paths, one through $X_1$ and one path
through~$X_2$.
\end{itemize}
%

\subsection{Two registers, covariates observed in only one of the
registers}\label{2part}\label{sec3.2}
In Section~\ref{theory2all} it is presumed that covariates are present
in both register $A$ as well as in register $B$.
Recently, it has been made possible to estimate the population size
making use of covariates that are only observed in one of the
registers [see Zwane
and van der Heijden (\citeyear{ZwaneEvanderHeijdenPGM2007}); for examples, see van
der Heijden, Zwane and Hessen (\citeyear{VanderHeijdenPGMZwaneEHessenD2009}), and Sutherland, Schwartz
and Rivest (\citeyear{SutherlandJMSchwarzCJRivestLP2007})].
A simple example illustrates the problem [see Panel 1 of Table
\ref{missing}] where covariate $X_1$ (Marital status) is only observed
in register $A$ (GBA) and covariate $X_2$ (Police region) is only
observed in register $B$ (HKS).
As a~result, $X_1$ is missing for those observations not in $A$ and
$X_2$ is missing for those observations not in $B$.
Zwane
and van der Heijden (\citeyear{ZwaneEvanderHeijdenPGM2007}) show that the missing observations
can be estimated using the EM algorithm under a missing-at-random
(MAR) assumption [Little
and Rubin (\citeyear{LittleRRubinD1987}),
Schafer (\citeyear{SchaferJ1997a,SchaferJ1997b})]
for the missing data process.
After EM, in a second step, the population size estimates are obtained
for each of the levels of $X_1$ and~$X_2$.

%
\begin{table}
\tabcolsep=0pt
\tablewidth=295pt
\caption{Covariate $X_1$ is only observed in register $A$
and $X_2$ is only observed in~$B$}\label{missing}
\begin{tabular*}{290pt}{@{\extracolsep{\fill}}lcccc@{}}
\multicolumn{5}{@{}l@{}}{\textit{Panel 1: Observed counts}}\\
\hline
&&\multicolumn{2}{c}{$\boldsymbol{A=1}$}\\[-5pt]
&&\multicolumn{2}{c}{\hrulefill} &\multicolumn{1}{c}{$\boldsymbol{A=2}$}\\
&&$\boldsymbol{X_1=1}$&$\boldsymbol{X_1=2}$&\multicolumn{1}{c@{}}{$\boldsymbol{X_1}$ \textbf{missing}}\\
\hline
$B=1$ &$X_2=1$ & 259 & 539 &  13,898  \\
&$X_2=2$ & 110 & 177 &  12,356  \\
[3pt]
$B=2$ & $X_2$ missing & \hphantom{0}91 & 164 & -- \\
\hline
\end{tabular*}\vspace*{6pt}
\begin{tabular*}{290pt}{@{\extracolsep{\fill}}lccccc@{}}
\multicolumn{6}{@{}l@{}}{\textit{Panel 2: Fitted values under
$[AX_2][BX_1][X_1X_2]$}}
\\
\hline
&&\multicolumn{2}{c}{$\boldsymbol{A=1}$}&\multicolumn{2}{c@{}}{$\boldsymbol{A=2}$}\\[-5pt]
&&\multicolumn{2}{c}{\hrulefill}&\multicolumn{2}{c@{}}{\hrulefill}\\
&&$\boldsymbol{X_1=1}$&$\boldsymbol{X_1=2}$&$\boldsymbol{X_1=1}$&$\boldsymbol{X_1=2}$ \\
\hline
$B=1$ &$X_2=1$ & 259.0 & 539.0 & 4510.8 & 9387.2 \\
&$X_2=2$ & 110.0 & 177.0 & 4735.8 & 7620.3 \\
[3pt]
$B=2$ & $X_2=1$& \hphantom{0}63.9 & 123.5 & 1112.4 & 2150.2 \\
& $X_2=2$ & \hphantom{0}27.1 & \hphantom{0}40.5 & 1167.9 & 1745.4 \\
\hline
\end{tabular*}
\vspace*{-2pt}
\end{table}

The number of observed cells is lower than in the standard situation.
For example, in Panel 1 of Table~\ref{missing} this number is $8$,
whereas it would have been $12$ if both $X_1$ and $X_2$ were observed
in both $A$ and $B$.
For this reason only a restricted set of loglinear models can be fit
to the observed data.
Zwane
and van der Heijden (\citeyear{ZwaneEvanderHeijdenPGM2007}) show that the most complicated model
is $[AX_2][BX_1][X_1X_2]$; note that the graph is similar to
the graph of $M_7$ in Figure~\ref{Figure2}, but $X_1$ and $X_2$ are
interchanged.
At first sight this model appears counter-intuitive, as one might
expect an interaction between variables $A$ and $X_1$, and between $B$
and $X_2$.
However, the parameter for the interaction between $A$ and $X_1$ (and
$B$ and $X_2$) cannot be identified, as the levels of $X_1$ do not vary
over individuals for which $A=2$.

This most complicated loglinear model $[AX_2][BX_1][X_1X_2]$ is
saturated, as the number of parameters is $8$ (namely, the general mean,
four main effect parameters and three interaction parameters) and
there are just $8$ observed values.
Consequently, these $8$ observed values are identical to the
corresponding $8$ fitted values.
The fitted values under this model are presented in Panel 2 of Table
\ref{missing}.
Note that, for example, the EM algorithm spreads out the observed
value 13,898 over the levels of $X_1$ into fitted values 4510.8 and
9387.2; note also that the ratio 4510.8/9387.2 of these fitted
values is identical to the ratio 259/539 of the observed values.

By comparison, when $X_1$ and $X_2$ are observed in both $A$ and $B$,
the saturated model is $M_{12}=[AX_1X_2][BX_1X_2]$.
This is a less restrictive model than the model
$[AX_2][BX_1][X_1X_2]$ and the difference is due to the MAR
assumption.

\begin{figure}

\includegraphics{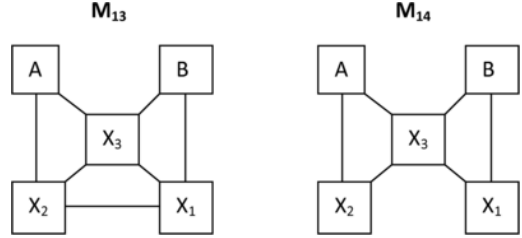}

\caption{Interaction graphs of loglinear models with partially
observed covariates.}\label{Figure3}
\end{figure}

We now consider the more general case when there are also covariates
observed in both $A$ and $B$.
Suppose that there is one covariate $X_1$ just observed in register
$A$, one covariate $X_2$ just observed in register $B$, and one
covariate $X_3$ observed in both registers.
The most complicated model is $M_{13}=[AX_2X_3][BX_1X_3][X_1X_2X_3]$,
with graph in Figure~\ref{Figure3}.
When $X_1$ and~$X_2$ are conditionally independent given $X_3$, the
model simplifies to $M_{14}=[AX_2X_3][BX_1X_3]$.
In $M_{14}$ there is only one short path, namely, $A-X_3-B$, and
neither covariate $X_1$ and $X_2$ is part of it.
Therefore, we can collapse the five-way table
$A\times B\times X_1\times X_2\times X_3$ over $X_1$ and $X_2$,
which illustrates Property 1.
We conclude that inclusion of covariates that are unique to\vadjust{\goodbreak}
specific
registers only modify the total population size estimate under the
model $M_{13}$, in which the covariates
just in $A$ are related to the covariates just in $B$.

Simplified situations exist when covariates $X_1$, $X_2$ or $X_3$ are
not available.
When $X_1$ is not available, $M_{13}$ reduces to model
$[AX_2X_3][BX_3]$, where the table $A\times B\times X_2\times
X_3$ is collapsible over $X_2$ because $X_2$ is not in the short path
$A-X_3-B$. Hence, to improve the total population size estimate,
covariates such
as $X_2$ are not useful unless $X_1$ both exists and is related to~$X_2$.
Similarly, when $X_2$ is not available, $M_{13}$ reduces to
$[AX_3][BX_1X_3]$ where the table is collapsible over $X_1$.
When the covariate $X_3$ is not available, $M_{13}$ reduces to model
$[AX_2][BX_1][X_1X_2]$, discussed earlier, where the covariates affect
the population size when $X_1$ is related to $X_2$.
If they are not related, the graph is similar to model $M_4$ and
collapsing the contingency table over both $X_1$ and $X_2$ does not
affect the total population size.

%
%

%
\begin{figure}[b]

\includegraphics{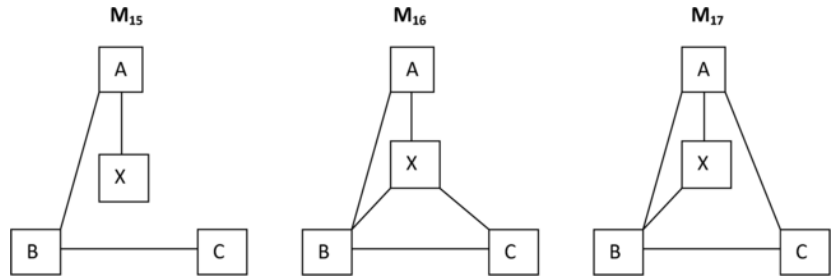}

\caption{Interaction graphs of loglinear models with three registers
and one covariate (see also next page).}\label{Figure4}
\end{figure}
%

\subsection{Three registers}\label{theory3}\label{sec3.3}
For completeness we give illustrative examples of the situation with
three or more registers
even though it is irrelevant for the data in Section~\ref{sec2}, where there
are only two.
For three registers $A$, $B$ and $C$ the contingency table
$A\times B\times{C}$ has one structural zero cell.
We consider how the Properties apply to the context of three registers
$A$, $B$ and~$C$, and with a single covariate $X$.
We discuss three models with their graphs displayed in Figure~\ref{Figure4}.

For model $M_{15}=[AX][AB][BC]$ the table $A\times B\times C\times
X$ is collapsible over covariate $X$, as it is not on any short path.
This illustrates Property 1. Property 2 is illustrated by the other
models where $A$ and $C$ are conditionally independent given $B$ and
$X$ is related to only one of the registers, namely, models
$[AB][BC][BX]$ and $[AB][BC][CX]$.

For model $M_{16}=[ABX][BCX]$ covariate $X$ is on the short path from~$A$ to $C$ and, therefore, the contingency table is not collapsible
over $X$. For model $M_{17}=[ABX][BC][AC]$ covariate $X$ is not on the
short path from $A$ to $B$, as the short path is $A - B$, and,
therefore, the contingency table is collapsible over $X$.

The maximal model $[ABX][BCX][ACX]$ is discussed at the end of
Appendix~\ref{append1}.

\section{Active and passive covariates}\label{actpass}\label{sec4}
In Section~\ref{theory} we discussed the result that marginalizing
over a covariate does not necessarily lead to a change in the
population size estimate.
Whether the population size estimate changes or not depends on the
loglinear models in the original and in the marginalized table.
We term a covariate \textit{active} if marginalizing over this
covariate leads to a different estimate in the reduced table, so that
this covariate plays an active role in determining the population
size; we call a covariate \textit{passive} if marginalizing leads to
an identical estimate in the reduced table.

As an example we discuss active and passive covariates referring to
Figure~\ref{Figure3}. We noted that in model $M_{13}$ the contingency table
is not collapsible over covariates $X_1$ and $X_2$, hence, they are
active covariates. On the other hand, in model $M_{14}$, by deleting
the edge between $X_1$ and $X_2$, the contingency table is collapsible
over $X_1$ and $X_2$, hence, they are passive covariates.

While passive covariates do not affect the size estimate, which
suggests that
they might be ignored, a possible use is the following.
A secondary objective of population size estimation is to provide
estimates of the size of subpopulations, or, equivalently, to break
down the population size in terms of given covariates.
This may well include passive covariates.
Describing a population breakdown in terms of passive covariates is an
elegant way to tackle this important practical problem.
This extends the approach of Zwane
and van der Heijden (\citeyear{ZwaneEvanderHeijdenPGM2007}) of using
register specific covariates in the population size estimation
problem.

Most registers have several covariates that are not common to other
registers, because the different registers are set up with different
purposes in mind.
An interesting data analytic approach is, therefore, first, to
determine a small number of active covariates, possibly of covariates
that are in both registers; and second, to set up a loglinear model
structured along the lines of model $M_{14}$, where several passive
covariates can be entered by extending~$X_1$ or $X_2$, and where these
covariates may or may not be register specific.
Passive covariates are helpful in breaking down the population size
under the assumption that the passive covariates of register $A$ are
independent of the passive covariates of register $B$ conditionally on
the active covariates.

We note that the introduction of many covariates may lead to sparse
contingency tables and hence to numerical problems due to empty
marginal cells in those margins that are fitted.
Consider, for example, a saturated model such as
$[AX_1X_2X_3][BX_1X_2X_3]$.
In this model the conditional odds ratios between $A$ and $B$ are 1.
However, when a zero count in one of the subtables of $X_1, X_2$ and
$X_3$ occurs for the levels of $A$ and of $B$, the estimate in this
subtable for the missing population is infinite. One way to solve this
is by setting higher order interaction parameters
equal to zero.

Another approach to tackle this numerical instability problem is as follows.
We start with an analysis using only active covariates, for example,
using the covariates observed in all registers in the saturated model.
We may monitor the usefulness of the model by checking the size of the
point estimate and its confidence interval.
If the usefulness is problematic (e.g., when the upper bound of
the parametric bootstrap confidence interval is infinite), we may make
the model more stable by choosing a more restrictive model.
One way to do this is by making a covariate passive.
For example, both in model $[AX_1X_2][BX_1X_2X_3]$ as well as in model
$[AX_1X_2X_3][BX_1X_2]$ the covariate $X_3$ is passive and both
models yield identical estimates and confidence intervals.
When one of these two model is chosen, its size may then be increased
by adding additional passive variables, such as variables that are
only observed in register $A$ or register $B$.
\section{Example}\label{sec5}
We now discuss the analysis of the data introduced in Section~\ref{sec2}.
To recapitulate, $A$ is inclusion in the municipal register GBA and
$B$ is inclusion in the police register HKS.
Covariates observed in both $A$ and~$B$ are $X_1$, gender,
$X_2$, age (four levels), and $X_3$, nationality (1~$=$ Iraqi; 2~$=$
Afghan; 3~$=$ Iranian).
Covariate $X_4$, marital status, is only observed in the
municipal register GBA.
Covariate $X_5$, police region where apprehended, with levels 1~$=$ in
one of the four largest cities of the Netherlands, and 2 $=$ elsewhere,
and is only observed in the police register HKS.

%
\begin{table}
\tabcolsep=0pt
\caption{Models fitted to example of variables $A, B, X_1$
to $X_5$, deviances, degrees of freedom, AIC's, estimated population
size and 95 percent confidence intervals}
\label{modexample}
\label{fit6}
\begin{tabular*}{\textwidth}{@{\extracolsep{\fill}}lcd{3.1}d{2.0}ccc@{}}
\hline
&\multicolumn{1}{c}{\textbf{Model}}&\multicolumn{1}{c}{\textbf{Deviance}}&\multicolumn{1}{c}{\textbf{df}}&\multicolumn{1}{c}{\textbf{AIC}}
&\multicolumn{1}{c}{\textbf{Pop. size}}&\multicolumn{1}{c@{}}{\textbf{CI}}\\
\hline
$N_1$& $[AX_1X_2X_3][BX_1X_2X_3]$ & 0 & 0 & 144.0 & 33,098.6 & 32,209--$\infty$\hphantom{468,}\\
$N_2$& $[AX_1X_2][BX_1X_2X_3]$ & 24.9& 16 & 136.8 & 33,504.1 & 32,480--35,468\\
$N_3$& $[AX_1X_2X_3][BX_1X_2]$ & 28.8& 16 & 140.7 & 33,504.1 & 32,480--35,468\\
[4pt]
$N_4$&$[AX_1X_2X_5][BX_1X_2X_3X_4]$ & 75.7&72&315.7& 33,504.1 & 32,480--35,468\\
$N_5$&$[AX_1X_2X_5][BX_1X_2X_3X_4][X_4X_5]$ & 75.7&71&317.7& 33,503.8 &
32,395--35,543\\
[4pt]
$N_6$&$[AX_1X_2X_3X_5][BX_1X_2X_4]$ & 523.8&72&763.7& 33,504.1 & 32,480--35,468\\
$N_7$&$[AX_1X_2X_3X_5][BX_1X_2X_4][X_4X_5]$ & 289.1&71&531.4& 33,510.9
& 32,363--35,432\\
\hline
\end{tabular*}
\vspace*{-3pt}
\end{table}

A first model is model $N_1=[AX_1X_2X_3][BX_1X_2X_3]$.
This is a saturated model.
For this model the estimate for the missed part of the population size
is $5504.6$, and the total population size is $33\mbox{,}098.6$.
However, the parametric bootstrap confidence interval [Buckland and
Garthwire (\citeyear{BucklandSGarthwireP1991})]  shows that we deal
with a solution that is numerically unstable, as the upper bound of the
$95$ percent confidence interval is infinite.
The instability of the model is a consequence of too many active
covariates, and a solution is to make covariate $X_3$ passive.
Two models in which $X_3$ is passive covariate are
$N_2=[AX_1X_2][BX_1X_2X_3]$ and $N_3=[AX_1X_2X_3][BX_1X_2]$.
For these models the population size estimate is
$33\mbox{,}504.1$ ($95$ percent CI is $32\mbox{,}481$--$35\mbox{,}469$).
Table~\ref{modexample} summarizes the results.

Models $N_2$ and $N_3$ are both candidates to be extended by including
marital status ($X_4$) or police region ($X_5$).
Note that $X_4$ is only observed in GBA ($A$) and $X_5$ is only
observed in HKS ($B$).
When $N_2$ is extended by adding~$X_4$ and $X_5$ as passive
variables, we get model $N_4[AX_1X_2X_5][BX_1X_2X_3X_4]$.
This model yields an identical estimate for the missed part of the
population, illustrating that in model $[AX_1X_2X_3X_5][BX_1X_2X_3X_4]$
the covariates~$X_4$ and $X_5$ are indeed passive.
With $72$ degrees of freedom and a deviance of~$75.7$ the fit is good.
The AIC is $315.7$.
We check whether it is better to make covariates $X_4$ and $X_5$
active and we do this by adding the interaction between the covariates
$X_4$ and $X_5$ to give model $N_5$.
The deviance of this model is identical and we conclude that $N_4$
is a better working model than~$N_5$.
We also extend $N_3$ by adding $X_4$ and $X_5$ as passive variables
giving $N_6$.
Note again that the estimate for the missed part of the population is
identical, however, the deviance is $523.8$ so the fit is worse.
Adding the interaction between $X_4$ and $X_5$ in $N_7$ helps as the
deviance goes to $289.1$, however, the deviance of $N_7$ is larger
than the deviance of $N_4$, so we choose~$N_4$ as the final model.

Out interest lies in the undocumented part of the population, that is,
in the people not registered in the GBA.
Table~\ref{margexample} shows the two-way margins of GBA with the
other variables estimated under $N_4$.
The estimates show that the undocumented population from Afghanistan,
Iraq and Iran are mostly not included in the police register HKS, are
more often male, between 25 and 50, from Afghanistan, unmarried and
mostly not staying in the four largest cities.\vspace*{-2pt}

%
%
\begin{table}
\caption{Estimates for GBA with each of the other variables
under model $N_4$}
\label{margexample}
\begin{tabular}{@{}lcccc@{}}
\hline
&\textbf{In HKS}& \textbf{Not in HKS}& \textbf{Male}& \textbf{Female}\\
\hline
In GBA & 1085.0 & 26,254.0 &15,855.0 & 11,484.0 \\
Not in GBA & \hphantom{0}255.0 & \hphantom{0,}5910.0 & \hphantom{0,}3874.7 & \hphantom{0,}2290.3 \\
[6pt]
&\textbf{15--25}& \textbf{25--35}& \textbf{35--50}& \textbf{50--64}\\
\hline
In GBA & 7234.0 & 8361.0 & 9185.0 & 2559.0\\
Not in GBA & 1292.2 & 2167.3 & 1925.9 & \hphantom{0}779.7\\
[6pt]
&\textbf{Afghan}&\textbf{Iraqi}&\textbf{Iranian}\\
\hline
In GBA & 12,818.8 & 8743.3 & 5776.8\\
Not in GBA & \hphantom{0,}2950.9 & 1914.5 & 1299.7\\
[6pt]
&\textbf{Unmarried}&\textbf{Married}&\textbf{4 large cities}& \textbf{Elsewhere}\\
\hline
In GBA & 14,698.2 & 12,640.8 & 9720.0 & 17,619.0\\
Not in GBA & \hphantom{0,}3302.3 & \hphantom{0,}2862.7 & 2182.6 & \hphantom{0,}3982.5\\
\hline
\end{tabular}
\end{table}

\section{Conclusion}\label{sec6}
We have demonstrated two closely related properties of loglinear models
in the context
of population size estimation.
First, under specific loglinear models marginalizing over covariates may
leave the population size estimate unchanged.
Second, different loglinear models fit to the same contingency table
may yield identical population size estimates.
This is worked out in detail for the case of two population registers
and illustrated for the three-register case.

Using the first property, we have introduced the notion of active and
passive covariates.
In a specific loglinear model, marginalizing over an active covariate
changes the population size estimate, while marginalizing over a~passive
variable leaves the population size estimate unchanged.
This idea can be particularly powerful in those situations where each
of the registers has unique covariates, but a description of the
full population in terms of these covariates is needed.
It may then be useful to introduce these register specific covariates
as passive covariates into a model such as $M_{14}$.
For example, if a loglinear model is proposed where the covariates
unique to register~$A$ are conditionally independent of the covariates
unique to register~$B$, then the full contingency tables is
collapsible over these covariates and, hence, these covariates are
passive.

Such a conditional independence assumption is strong, yet in many
data sets there may not be enough power to test its correctness.
It is demonstrated that a direct relation between the passive
covariates of register~$A$ and those in $B$ can only be assessed among
those individuals that are in both register~$A$ and $B$.
If there is {overlap} between register $A$ and $B$, with relatively
many individuals in both $A$ and $B$, the relationship between the
passive covariates of $A$ and $B$ can easily be assessed; conversely,
if the overlap is {small}, there is little power to establish whether
or not this relation should be included in the model.

This new methodology should be of use for estimating the missing
population due to undercoverage in the 2011 Census of the Netherlands
where the size of the total population can be estimated by application
of loglinear models.
It could also be applied to countries that use register information to
estimate the undercoverage of their Population Register as well as to
countries which use traditional methods.
The use of passive covariates gives insight into which characteristics
individuals have that are not covered by the Census and thereby
illuminate the bias due to the undercoverage.

In the \hyperref[sec1]{Introduction} we mentioned latent variable models that take
heterogeneity of inclusion probabilities into account. For this purpose
both \citet{FienbergSJohnsonMJunkerB1999} as well as in Bartolucci
and Forcina (\citeyear{BartolucciFForcinaA2001})
proposed generalizations of the so-called Rasch model. It is beyond the
scope of this paper to study collapsibility properties for their models in
the presence of covariates. However, it is interesting to note that one
important specific form of the Rasch model, the so-called extended Rasch
model, is mathematically equivalent to the loglinear model that includes
three two-factor interactions that are identical and a three-factor
interaction [see Hessen
(\citeyear{HessenDJ2011}); this loglinear
model is also used in
IWGDMF (\citeyear{IWGfDMaF1995}), where it is referred to as a heterogeneity
model]. Collapsibility properties of this loglinear model can be
studied using the perspective presented in this paper.\looseness=1

\begin{appendix}
\section{Identification of equivalent models}\label{sec7}\label{append1}
We establish which models listed in Figures~\ref{Figure1}--\ref{Figure4} have the same
estimates, and which do not, by showing that models for population
size estimation are model collapsible onto two margins; and by
demonstrating how the short path criterion identifies noninvariance
of population size estimates. Our method is to apply the Asmussen and
Edwards (\citeyear{AsmussenSEdwardsD1983}) criterion to the population size estimation model which
contains structural zeros.
\subsection{Model collapsibility}\label{sec7.1}
First we recall the model collapsibility condition of
Asmussen
and Edwards (\citeyear{AsmussenSEdwardsD1983}).
Consider a table classified by two sets of factors $Y$ and $Z$, so
that the saturated model is $[YZ]$, and maximum likelihood estimation
under product multinomial sampling.
The authors give conditions on the hierarchical loglinear model
$M\subset[YZ]$ under which
%
\begin{eqnarray}\label{eq:modelcollap}
\hat{p}^{N}_{Y}(y) = \sum_{z} \hat{p}^{M}_{YZ}(y,z),
\end{eqnarray}
where the right-hand side (RHS) is the margin of the MLE under the
model~$M$ for the full table, while the LHS is the MLE under the
restricted model~$N$ for the margin obtained by deleting terms in $Z$
from each generator of~$M$.
Their Theorem~2.3 states that~$M$ is (model) collapsible onto the
margin~$Y$,
that is, (\ref{eq:modelcollap}) holds, if and only if the boundary of every
connected component of $Z$ is contained in a generator of $M$.
A corollary to this result is that estimates computed under $N$
have the same sampling distribution as those under $M$, and hence
the same confidence intervals.

Implicit in their derivation is that the space on which the
table is defined is a Cartesian product of the factors.
We argue that the population size estimation model cannot be defined on
a Cartesian product of registers, for in our context if $p$
were
defined on
$\calA\times\calB\times\calX$ with $\calA, \calB=\{1,2\}$, then we
require $p(2,2,x)=0$ to reflect a structural zero.
If so, the maximal loglinear model would be $M = [ABX]$ with a three factor
interaction, as $\log p$ contains the interaction term
$\lambda_{ABX}(2,2,x)=-\infty$.
Furthermore, application of model collapsibility suggests $M = [ABX]$
is model collapsible onto $[AB]$, which may be shown by counterexample
to be false.

\subsection{Models for population size estimation}\label{sec7.2}\label{append2}
For population size estimation the appropriate sample space $\calS$
for two registers is
\begin{eqnarray*}\label{eq:}
\calS=\{(a,b); (a,b)=(1,1),(1,2),(2,1)\},
\end{eqnarray*}
as $(2,2)$ cannot be observed, and the sample space for the whole
survey is $\calS\times \mathcal{X}$, where $\mathcal{X}$ is the
Cartesian product of the discrete spaces for the covariates.
Any loglinear model $M$ with probability mass function $p^{M}_{SX}$ is
defined and fitted on this space.
The loglinear expansion of $\log p^{M}_{SX}(a,b,x)$ under
the maximal model $M=[AX][BX]$ is
%
\begin{eqnarray}\label{eq:maxmodel}
\lambda+
\lambda_{A}(a)+
\lambda_{B}(b)+
\lambda_{X}(x)+
\lambda_{AX}(a,x)+
\lambda_{BX}(b,x)
\end{eqnarray}
for $(a,b,x)\in\calS\times\calX$.
The $\lambda$ parameters satisfy corner point constraints to ensure
identifiability, but are otherwise arbitrary.
This is an instance of a~hierarchical loglinear model; an equivalent
parameterization is to write the highest order main effect as
$\lambda_{SX}(s,x)$, but this obscures the submodels of interest.
The register $A$ taking values in $\calA$ defines the marginal
probability~$p^{M}_{AX}$ of $p^{M}_{SX}$, similarly $p^{M}_{BX}$.

Asmussen
and Edwards (\citeyear{AsmussenSEdwardsD1983}) define the
interaction graph to be the graph with a node for each factor
classifying the table and an edge between two nodes if there is a
generator in the model containing both.
Consequently, the graphs in Figures~\ref{Figure1}--\ref{Figure4} are the interaction graphs of
particular population size models.
The interaction graph of $M=[AX][BX]$ is that of $M_{3}$ in Figure~\ref{Figure1} with $X$ replacing~$X_{1}$.

These graphs cannot be interpreted as conditional independence graphs
in which the missing edge between $A$ and $B$ leads to the statement
$A\indep B|X$, as this is false on the restricted space
$\calS\times \mathcal{X}$; for instance, if $X$ is empty, and
$M=[A][B]$, then ${P}(A=1,B=1)\neq {p_{A}}(1){p_{B}}(1)$.
However, conditional independence interpretations between a register
and covariates, and between two covariates are possible.

With the population size estimation model at (\ref{eq:maxmodel}) defined on the right
space, $\calS\times\calX$, we can now employ model collapsibility to
show this model is collapsible onto two margins.

\subsection{Model collapsibility for population size estimation}\label{sec7.3}
Our first result is that the maximal population size model in
(\ref{eq:maxmodel}) is model collapsible onto its two margins $[AX]$
and $[BX]$.
Standard arguments show the sufficient statistics are $n_{AX}(a,x)$
and
$n_{BX}(b,x)$, where $n$ is the frequency function of the observations
over the table.
Under this model the MLEs satisfy
$\hat{p}^{M}_{AX}=n_{AX}(a,x)/n_{\varnothing}$ and
$\hat{p}^{M}_{BX}=n_{BX}(b,x)/n_{\varnothing}$;
and these margins determine the full table $\hat{p}^{M}_{SX}$.
To apply (\ref{eq:modelcollap}) when marginalizing over $B$, note the
boundary of $\{A,B,X\}\exc B$ in the interaction graph is $\{A,X\}$,
and that these factors are both contained in a single generator of
$M$, namely, $[AX]$.
Similarly for marginalizing over $A$ so that the model is collapsible
onto the two margins, and
%
\begin{eqnarray}\label{eq:collmargins}\quad
  \hat{p}^{M}_{AX}(a,x) = \sum_{b} \hat
{p}^{M}_{SX}(a,b,x),\qquad
\hat{p}^{M}_{BX}(b,x) = \sum_{a} \hat{p}^{M}_{SX}(a,b,x).
\end{eqnarray}
%
\subsection{Population size estimation invariance}\label{sec7.4}
We define population size estimation invariance, and show it depends
on the model collapsibility of the population size model onto two
margins, both containing one register and the covariates. Examples
are given.

A population size estimate is made by extending the fitted
probability~$p^{M}_{SX}$ on $\calS\times \calX$ to $\pi^{M}$ defined on the
Cartesian product space $\calA\times \calB\times \calX$, by the
conditional independence statement
\begin{eqnarray*}\label{eq:}
\pi^{M}(a,b,x) = p^{M}_{AX}(a,x)  p^{M}_{BX}(b,x)  / p^{M}_{X}(x)
\jfor(a,b,x)\in\calA\times \calB\times \calX.
\end{eqnarray*}
Under the measure $\pi$ the interaction graphs in Figures
\ref{Figure1}--\ref{Figure4} now have
conditional independence interpretations.

The fitted values for $\hat{\pi}^{M}$ are computed from the fitted
values $\hat{p}^{M}_{AX}$ and~$\hat{p}^{M}_{BX}$ which are obtained from
$\hat{p}^{M}(a,b,x)$ fitted on $\calS^{2}\times \calX$ at
(\ref{eq:collmargins}).
The population size estimate is $n_{\varnothing}(1+\hat{\pi
}^{M}(2,2))$, where
%
\begin{eqnarray}\label{eq:popsize}
\hat{\pi}^{M}(a,b) &=& \sum_{x}
\hat{p}^{M}_{AX}(a,x)  \hat{p}^{M}_{BX}(b,x)  / \hat{p}^{M}_{X}(x).
\end{eqnarray}
Two loglinear models $M$ and $N$ have identical population size
estimates whenever $\hat{\pi}^{M}(a,b)=\hat{\pi}^{N}(a,b)$ for all
$(a,b)\in\calA\times \calB$.
So because of (\ref{eq:popsize}) the condition for invariance devolves
to model collapsibility of ${M}$ on $\calA\times\calX$ and on
$\calB\times\calX$.

We illustrate population size estimation invariance by showing that
certain models for $\pi$ displayed in the figures above have identical
estimates.
The first example shows
the model $M_{2}=[A][BX_{1}]$ in Figure~\ref{Figure1} is collapsible
on $X_{1}$ to $M_{0}=[A][B]$, and so produces identical population
size estimates.
From (\ref{eq:popsize})
\begin{eqnarray*}\label{eq:}
\hat{\pi}^{(2)}(a,b)
&=& \sum_{x_{1}}\hat{p}^{(2)}_{A}(a)  \hat{p}^{(2)}_{BX_{1}}(b,x_{1}),
\end{eqnarray*}
by the independence of $A$ and $X_{1}$ under $M_{2}$.
By the model collapsibility of~$[BX_{1}]$ over $X_{1}$,
\begin{eqnarray*}\label{eq:}
\hat{\pi}^{(2)}(a,b) = \hat{p}^{(2)}_{A}(a)  \sum_{x_{1}}\hat
{p}^{(2)}_{BX_{1}}(b,x_{1}) = \hat{p}^{(0)}_{A}(a)  \hat{p}^{(0)}_{B}(b),
\end{eqnarray*}
which is just $\hat{\pi}^{(0)}(a,b)$ as required.

The second example is to show the model
$M_{11}=[AX_{1}][BX_{1}X_{2}]$ in Figure~\ref{Figure1}
is collapsible on $X_{2}$ to $M_{3}=[AX_{1}][BX_{1}]$, and so
produces identical population size
estimates.
From (\ref{eq:popsize}), using the independence
$A$ and $X_{2}$ given $B,X_{1}$
under $M_{11}$,
\begin{eqnarray*}\label{eq:}
\hat{\pi}^{(11)}(a,b)
&=& \sum_{x_{1},x_{2}}\hat{p}^{(11)}_{AX_{1}}(a,x_{1})  \hat
{p}^{(11)}_{BX_{1}X_{2}}(b,x_{1},x_{2})   /\hat
{p}^{(11)}_{X_{1}}(x_{1}),\\[-2pt]
&=&\sum_{x_{1}} \hat{p}^{(11)}_{AX_{1}}(a,x_{1})  /\hat
{p}^{(11)}_{X_{1}}(x_{1})
\sum_{x_{2}}\hat{p}^{(11)}_{BX_{1}X_{2}}(b,x_{1},x_{2}) \\[-2pt]
&=&\sum_{x_{1}} \hat{p}^{(3)}_{AX_{1}}(a,x_{1})  \hat
{p}^{(3)}_{BX_{1}}(b,x_{1})
  /\hat{p}^{(3)}_{X_{1}}(x_{1}),
\end{eqnarray*}
by the collapsibility of each of the three components in the expression
and equals $\hat{\pi}^{(3)}(a,b)$ by definition.

\subsection{Short path criterion for population size invariance}\label{sec7.5}
We demonstrate how the short path criterion identifies noninvariance
in the context of an example attempting to argue that $M_{7}$ produces
identical estimates to~$M_{3}$.

First consider the population size estimate from $M_{7}$:
\begin{eqnarray*}\label{eq:}
\hat{\pi}^{(7)}(a,b)
&=&\sum_{x_{1},x_{2}}
\hat{p}^{(7)}_{AX_{1}X_{2}}(a,x_{1},x_{2})
\hat{p}^{(7)}_{BX_{1}X_{2}}(b,x_{1},x_{2})/
\hat{p}^{(7)}_{X_{1}X_{2}}(x_{1},x_{2}).
\end{eqnarray*}
Using the two independences under $M_{7}$,
\begin{eqnarray*}\label{eq:}
\hat{\pi}^{(7)}(a,b) &=&\sum_{x_{1},x_{2}}
\hat{p}^{(7)}_{AX_{1}}(a,x_{1})
\hat{p}^{(7)}_{BX_{2}}(b,x_{2})
\hat{p}^{(7)}_{X_{1}X_{2}}(x_{1},x_{2})/\hat
{p}^{(7)}_{X_{1}}(x_{1})\hat{p}^{(7)}_{X_{2}}(x_{2})
\\[-2pt]
&=& \sum_{x_{1}}\hat{p}^{(7)}_{AX_{1}}(a,x_{1})/\hat{p}^{(7)}_{X_{1}}(x_{1})
\sum_{x_{2}}\hat{p}^{(7)}_{BX_{2}}(b,x_{2})
\hat{p}^{(7)}_{X_{1}X_{2}}(x_{1},x_{2})/
\hat{p}^{(7)}_{X_{2}}(x_{2}).
\end{eqnarray*}
While model collapsibility implies $\hat{p}^{(7)}_{AX_{1}}(a,x_{1})=
\hat{p}^{(3)}_{AX_{1}}(a,x_{1})$, simple counter examples show
$\hat{p}^{(3)}_{BX_{1}}(b,x_{1}) \neq\sum_{x_{2}}\hat
{p}^{(7)}_{BX_{2}}(b,x_{2})
\hat{p}^{(7)}_{X_{1}X_{2}}(x_{1},x_{2})/
\hat{p}^{(7)}_{X_{2}}(x_{2})$.
Here $X_2$ is on a short path from $A$ to $B$ and the population size
estimates are not invariant to marginalizing over $X_2$.

The last model we consider is the maximal model for three registers
$A$,~$B$ and $C$ and covariate $X$, that is, $[ABX][ACX][BCX]$. It is
collapsible over $A$, or $B$, or $C$, but it is not collapsible
over
$X$. Of course, population size estimates are not invariant to
collapsing over $A$ even though $[ABX][ACX][BCX]$ is model collapsible
over $A$, showing that
population size invariance is not equivalent to model collapsibility.

\section{Estimation}\label{sec8}

Estimation of the missing count can be done as follows. We first
discuss the case that there is no covariate. Let $A$ and $B$ have
levels $a, b = 1,2$, for ``registered'' and ``not registered.'' We denote
observed frequencies by~$n_{ab}$ with $(a, b) = (2, 2)$ missing.
Expected frequencies are denoted by~$m_{ab}$ and fitted values by $\hat
{m}_{ab}$. For the three cells $(a, b)$ with $(a, b) \ne(2, 2)$ we
define a~loglinear independence model as log $m_{ab} = \lambda+
\lambda_{A}(a) + \lambda_{B}(b)$ with $\lambda_{A}(2) = \lambda
_{B}(b) = 0$. Then, after fitting the loglinear model, the missing
count $m_{22}$ is found as $\hat{m}_{22} = \exp (\hat{\lambda})$.

In the presence of a covariate $X$ with levels $x = 1,2$, the observed
counts are $n_{abx}$ with $(a, b, x) = (2, 2, x)$ missing. A saturated
loglinear model for the six observed counts is log $m_{abx} = \lambda+
\lambda_{A}(a) + \lambda_{B}(b) + \lambda_{X}(b) + \lambda_{AX}(ax)
+ \lambda_{BX}(bx)$ with $\lambda_{A}(2) = \lambda_{B}(2) = \lambda
_{X}(2) = 0$. Then, after fitting a saturated or restricted loglinear
model to the six observed counts, the missing counts are found as $\hat
{m}_{221}  = \exp   (\hat{\lambda} + \hat{\lambda}_{X}(1)
)$ and $\hat{m}_{222} = \exp \hat{\lambda}$. This generalizes in
a~natural way to the situation that there are more registers, that
covariates have more than two levels and more covariates.

Extra information is needed for the models in Section~\ref{2part},
where covariates are observed in only one of the registers. We follow
the explanation in Zwane
and van der Heijden (\citeyear{ZwaneEvanderHeijdenPGM2007}). The approach taken
to analyze such data (data with partly available covariates) is to
identify the problem as a missing information problem, and then use the
EM algorithm to obtain maximum likelihood estimates.

The EM algorithm is an iterative procedure with two steps, namely, the
expectation and maximization step. The EM algorithm starts with initial
values for the probabilities to be estimated. Initial values have to be
at the interior of the parameter space (i.e., not equal to zero), for
example, form a uniform table, in which all the elements are equal. In
the $t$th E-step, we compute the expected loglikelihood of the complete
data conditional on the available data under the values of the
parameters in that iteration. In the $t$th M-step, a loglinear model is
fitted to the completed data, with the missing cells corresponding to
$(a, b) = (2, 2)$ denoted as structurally zero. The fitted
probabilities under the loglinear model fitted in the M-step are then
used in the E-step of the ($t + 1$) iteration, to derive updates for
the completed data.

Cycling between the E-step and the M-step goes on until convergence. At
each iteration the likelihood increases. Convergence to a local maximum
or a saddle point is guaranteed. Schafer [(\citeyear{SchaferJ1997a}), pages 51--55] states
that, in well-behaved problems (i.e., problems with not too many
missing entries and
not too many parameters), the likelihood function will be unimodal and
concave on the entire parameter space, in which case EM converges to
the unique maximum likelihood estimate from any starting value. Thus
far, we have never encountered examples where multiple maxima exist,
and a~typical way to investigate the presence of multiple maxima is by
trying out different starting values.

After convergence, the fit is assessed using the observed elements only
(e.g., for Table~\ref{missing} there are only 8 observed
elements, whereas in the completed table, excluding the structural zero
cells, there are 12 elements). Degrees of freedom are determined using
the number of observed elements minus the number of fitted parameters.

The values for the missing cells corresponding to $(a, b) = (2, 2)$ are
assessed using the method that we described above.

We use parametric bootstrap confidence intervals because they provide
a~simple way to find the confidence intervals when the contingency
table is not fully observed. To compute the bootstrapped confidence
intervals for a~specific loglinear model, we need to first compute the
population size under this model
and the probabilities on the completed data under this model,
that is, by including the cells that cannot be observed by design.
A~first multinomial sample is drawn given these parameters, and the
sample is then reformatted to be identical to the observed data. The
specific loglinear model used is then fitted to the resulting data,
resulting in the first bootstrap sample estimate of the population
size. If $K$ bootstrap samples are needed, then this is repeated $K$
times. By ordering the $K$ bootstrap population size estimates, a
confidence interval can be constructed.
\end{appendix}

\begin{supplement}
\stitle{Estimation in R}
\slink[doi]{10.1214/12-AOAS536SUPP} 
\slink[url]{http://lib.stat.cmu.edu/aoas/536/supplement.pdf}
\sdatatype{.pdf}
\sdescription{We make use of the CAT-procedure in R
(Meng
and Rubin (\citeyear{MengXLRubinDB1991}); Schafer
[(\citeyear{SchaferJ1997a}), Chapters 7 and 8], (\citeyear{SchaferJ1997b})).
The CAT-procedure is a routine for the analysis of categorical variable
data sets with missing values.
We describe our application of this procedure in detail in the
supplemental article [van
der Heijden et al. (\citeyear{vanderHeijdenPGMWhittakerJetal2012})].}
\end{supplement}

%

\printaddresses

\end{document}